\documentclass[aps,prb,twocolumn,superscriptaddress,showpacs]{revtex4}
\usepackage{amsmath}
\usepackage{bm}
\usepackage{graphicx}

\begin{document}

\title{Fano and Dicke effects and spin polarization in a double Rashba-ring system side
coupled to a quantum wire}
\author{V.M. Apel}
\affiliation{Departamento de F\'{\i}sica, Universidad Cat\'{o}lica
del Norte, Casilla 1280, Antofagasta, Chile}
\author{P. A. Orellana}
\affiliation{Departamento de F\'{\i}sica, Universidad Cat\'{o}lica del Norte, Casilla
1280, Antofagasta, Chile}
\author{M. Pacheco}
\affiliation{Departamento de F\'{\i}sica, Universidad T\'ecnica F. Santa Maria, Casilla
postal 110 V, Valpara\'iso, Chile}

\begin{abstract}
The electronic transport in a system of two quantum rings
side-coupled to a quantum wire is studied via a single-band
tunneling tight-binding Hamiltonian. We derived analytical
expressions for the conductance and spin polarization when the
rings are threaded by magnetic fluxes with Rashba spin-orbit
interaction. We show that by using the Fano and Dicke effects this
system can be used as an efficient spin-filter even for small spin
orbit interaction and small values of magnetic flux. We compare
the spin-dependent polarization of this design and the
polarization obtained with one ring side coupled to a quantum
ring. As a main result, we find better spin polarization
capabilities as compared to the one ring design
\end{abstract}

\maketitle

\section{Introduction}

Electronic transport through quantum rings structures has become the subject
of active research during the last years. Interesting quantum interference
phenomena have been predicted and measured in these mesoscopic systems in
presence of a magnetic flux, such as the Aharonov-Bohm oscillations in the
conductance, persistent currents\cite{Chandra,Mailly,Keyser} and Fano
antiresonances \cite{damato,pedro}.

Recently, there has been much interest in understanding the manner
in which the unique properties of nanostructures may be exploited
in spintronic devices, which utilize the spin degree of freedom of
the electron as the basis of their
operation~\cite{Datta,Song,Folk,Mireles,Mireles2,Berciu}. A
natural feature of these devices is the direct connection between
their conductance and their quantum-mechanical transmission
properties, which may allow their use as an all-electrical means
for generating and detecting spin
polarized distributions of carriers. For instance, recently Son et al.\cite%
{Song} described how a spin filter may be realized in open-quantum dot
system, by exploiting the Fano resonances that occur in their transmission.
In a quantum dot in which the spin degeneracy of carrier is lifted, they
showed that the Fano effect may be used as an effective means to generate
spin polarization of transmitted carriers, and that electrical detection of
the resulting polarization should be possible. This idea was extended to
side attached quantum rings. In Ref.(\onlinecite{Shelykh}) Shelykh et. al.
analyze the conductance of the Aharonov-Bohm (AB), one-dimensional quantum
ring touching a quantum wire. They found that the period of the AB
oscillations strongly depends on the chemical potential and the Rashba
coupling parameter. The dependence of the conductance on the carrier's
energy reveals the Fano antiresonances. On the other hand, Bruder et. al.%
\cite{bruder} introduce a spin filter based on spin-resolved Fano resonances
due to spin-split levels in a quantum ring side coupled to a quantum wire.
Spin-orbit coupling inside the quantum ring, together with external magnetic
fields, induces spin splitting, and the Fano resonances due to the
spin-split levels result in perfect or considerable suppression of the
transport of either spin direction. They found that the Coulomb interaction
in the quantum ring enhances the spin-filter operation by widening the
separation between dips in the conductance for different spins and by
allowing perfect blocking for one spin direction and perfect transmission
for the other.

\begin{figure}[h!]
\centerline{\includegraphics[width=7cm,angle=0,scale=0.7]{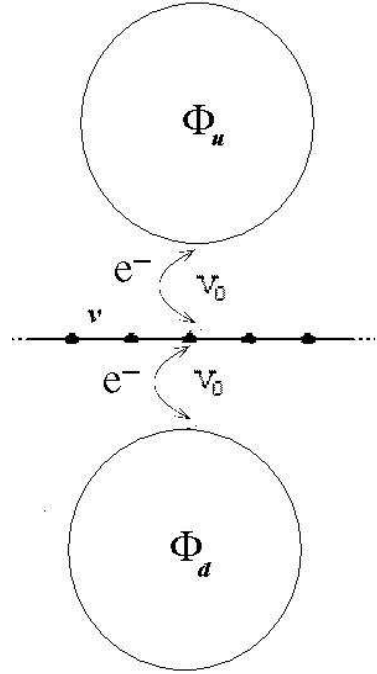}}
\caption{Schematic view of the two quantum ring attached to quantum wire.}
\label{fig1}
\end{figure}
In this paper we study the two ring side-coupled to a quantum wire
in presence of magnetic flux and Rashba spin-orbit interaction, as
shown schematically in Fig.~\ref{fig1}. In a previous paper (ref
\onlinecite{pedro2}) we investigate the conductance and the
persistent current of two mesoscopic quantum ring attached to a
perfect quantum wire in presence of a magnetic field. We show that
the system develops an oscillating band with resonances (perfect
transmission) and antiresonances (perfect reflection). In
addition, we found persistent current magnification in the rings
due to the Dicke effect in the rings when the magnetic flux
difference is an integer number of the quantum of flux. The Dicke
effect in optics takes place in the spontaneous emission of a pair
of atoms radiating a photon with a wave length much larger than
the separation between them. \cite{dicke} The luminescence
spectrum is characterized by a narrow and a broad peak, associated
with long and short-lived states, respectively. Now, we show that
by using the Fano and Dicke effects this system can be used as an
efficient spin-filter even for small spin orbit interaction and
small values of magnetic flux. We find that the spin-polarization
dependence for this system is much more sensitive to magnetic flux
and spin-orbit interaction than the case with only one ring
side-coupled to the quantum wire.

\section{Model}

In the presence of Rashba the spin-orbit coupling and magnetic
flux $\Phi _{AB}$, the Hamiltonian for an isolated one-dimensional
rings reads \cite{Meijer},

\begin{equation}
H=\hbar \Omega \left[ \left( -i\frac{\partial }{\partial \varphi }-\frac{%
\Phi _{AB}}{\Phi _{0}}+\frac{\omega _{so}}{\Omega }\sigma _{r}(\varphi
)\right) ^{2}-\frac{\omega _{so}^{2}}{4\Omega ^{2}}\right]
\end{equation}

where,
\begin{equation*}
\sigma _{r}(\varphi )=\cos (\varphi )\sigma _{x}+\sin (\varphi )\sigma _{y}
\end{equation*}%
where $\sigma _{x}$ , $\sigma _{y}$ and $\sigma _{z}$ are the Pauli
matrices. The parameter $\hbar \Omega =\frac{\hbar ^{2}}{2ma^{2}}$and $%
\omega _{so}=\frac{\alpha _{so}}{\hbar a}$ is the frequency
associated to the SO\ coupling. The spin-orbit coupling constant
$\alpha_{so}$ depends implicitly on the strength of the surface
electric field \cite{luo}. The energy spectrum of the above
Hamiltonian is given by,

\begin{equation}
\varepsilon _{\mu n}=\hbar \Omega \left[ \left( n-\phi _{AB}+\frac{1}{2}-\mu
\frac{1}{2\cos \theta }\right) ^{2}-\frac{1}{4}\tan ^{2}\theta \right]
\end{equation}

\noindent where $\theta =-\arctan (\omega _{so}/\Omega )$ and $\phi _{AB}=$ $%
\frac{\Phi _{AB}}{\Phi _{0}}$, the Aharonov-Bohm phase.

The eigenstates are given by the following wave functions,

\begin{equation*}
\Psi _{n}^{+}(\varphi )=e^{in\varphi }\left(
\begin{array}{c}
\cos (\frac{\theta }{2}) \\
e^{i\varphi }\sin (\frac{\theta }{2})%
\end{array}%
\right)
\end{equation*}

\begin{equation*}
\Psi _{n}^{-}(\varphi )=e^{in\varphi }\left(
\begin{array}{c}
\sin (\frac{\theta }{2}) \\
-e^{i\varphi }\cos (\frac{\theta }{2})%
\end{array}%
\right)
\end{equation*}

The second quantization form of the quantum wire-quantum ring
device with a magnetic flux and spin-orbit interaction can be
written as,

\begin{widetext}
\begin{equation}
H_{T}=\sum_{i\mu }\varepsilon _{i}c_{\mu ,i}^{\dag }c_{\mu
,i}+v\sum_{\left\langle ij\right\rangle \mu }\left( c_{\mu
,i}^{\dag }c_{\mu ,j}+h.c\right) +\sum_{\alpha ,n,\mu }\varepsilon
_{\mu ,n}^{\alpha }d_{\mu ,n}^{\alpha \dag }d_{\mu ,n}^{\alpha
}+V_{0}\sum_{\mu ,n,\alpha }(d_{\mu ,n}^{\alpha \dag }c_{\mu
0}+h.c)
\end{equation}
\end{widetext}

The operator $c_{j\mu }^{\dag }$ creates an electron in the site $j$ of the
wire and with spin index $\mu $, $d_{n\mu }^{\alpha \dag }$ creates an
electron in the level $n$ of the ring $\alpha $ and with spin index $\mu$.
The wire site-energy is assumed equal to zero and the hopping energies for
wire and rings are taken to be equal to $v$, whereas $V_{0}$ couples both
systems.

Within the described model the conductance can be calculated by means of a
Dyson equation for the Green's function.

\begin{equation}
G_{\mu 0}^{\alpha }=\frac{i}{2v\sqrt{1-\frac{\omega ^{2}}{4v^{2}}}}\frac{1}{%
1-i\gamma \sum\limits_{\beta }A_{\mu }^{\beta }(\omega )}
\end{equation}

where $\gamma =\frac{V_{0}^{2}}{2v\sqrt{1-\frac{\omega ^{2}}{4v^{2}}}}$ and

\begin{equation}
A_{\mu }^{\alpha }(\omega )=\sum_{n=-\infty }^{\infty }g_{n\mu }^{\alpha
}=\sum_{n=-\infty }^{\infty }\frac{1}{\omega -\varepsilon _{\mu n}^{\alpha }}
\end{equation}
and,
\begin{equation}
g_{n\mu }^{\alpha }=\frac{1}{\omega -\varepsilon _{\mu n}^{\alpha }}.
\end{equation}

\noindent Where $g_{n\mu }^{\alpha}$ is the Green's function of the isolated
ring $\alpha$.

The conductance of the system can be calculated using the Landauer
formula.

\begin{equation}
\mathcal{G}_{\mu }=\frac{e^{2}}{h}T_{\mu }\left( \omega =E_{F}\right)
\end{equation}

\noindent where $T_{\mu }$ is the probability transmission. In the
linear response approach it can be written in term of the Green's
function of the contact by:

\begin{equation}
T_{\mu }\left( \omega \right) =\Gamma \left( \omega \right) \Im m \left[
G_{\mu 0}^{\alpha }\left( \omega \right) \right] =\frac{1}{1+\gamma ^{2}%
\left[ \sum\limits_{\beta }A_{\mu }^{\beta }(\omega )\right] ^{2}},
\end{equation}

\noindent where $\Gamma(\omega)=2v\sqrt{1-\frac{\omega^{2}}{4v^{2}}}$.

\bigskip

Following ref.~\onlinecite{Song} we introduce the weighted spin
polarization as

\begin{equation}
P_{\mu }=\frac{\left\vert T_{+}-T_{-}\right\vert }{\left\vert
T_{+}+T_{-}\right\vert }\,T_{\mu }\ ,\qquad \mu =\pm .  \label{wsp}
\end{equation}

Notice that this definition takes into account not only the relative
fraction of one of the spins, but also the contribution of those spins to
the electric current. In other words, we will require that not only the
first term of the right-hand side of~(\ref{wsp}) to be of order of unity,
but also the transmission probability $T_{\mu}$.

\section{Results}

In what follow we present results for the conductance and spin polarization
for a double ring system of radius $a=120nm$ , coupled each other through a
quantum-wire. For this radius the energy $\hbar \Omega=40\mu eV$. We
consider only energies near of the center of the band therefore we consider
the tunneling coupling as a constant. Then we set the tunneling coupling $%
\gamma=16\mu eV$.

By using the results given is ref.[\onlinecite{wunsch}] $A_{\mu
}^{\beta }(\omega )$ can be evaluated analytically,

\begin{eqnarray*}
A_{\mu }^{\alpha }(\omega ) &=&\frac{2\pi ^{2}}{\hbar \Omega z}\frac{\sin (z)%
}{\cos (2\pi \phi _{\mu }^{\alpha })-\cos (z)} \\
z &=&\pi \left( \frac{4\omega }{\hbar \Omega }+\frac{\omega _{so}^{2}}{%
\Omega ^{2}}\right) ^{1/2}
\end{eqnarray*}%
\noindent where, $\phi _{\mu }^{\alpha }=\phi _{AB}^{\alpha }+\frac{1}{2}%
-\mu \frac{1}{2\cos \theta },$ is the net phase for the $\alpha $-ring.
Then, we can obtain an analytical expression for the conductance,

\bigskip

\begin{widetext}
\begin{equation}
\mathcal{G}_{\mu}(\omega)=\frac{e^{2}}{h}\frac{\left[ \left( \cos
(2\pi \phi _{\mu }^{u})-\cos
(z)\right) \left( \cos (2\pi \phi _{\mu }^{d})-\cos (z)\right) \right] ^{2}}{%
\left[ \left( \cos (2\pi \phi _{\mu }^{u})-\cos (z)\right) \left(
\cos (2\pi \phi _{\mu }^{d})-\cos (z)\right) \right] ^{2}+\beta
^{2}\left[ \cos (2\pi \phi _{\mu }^{u})+\cos (2\pi \phi _{\mu
}^{d})-2\cos (z)\right] ^{2}}.\label{conduc1}
\end{equation}%
\end{widetext}

\noindent with $\beta =\left( \gamma 2\pi ^{2}/\hbar \Omega \right) \left(
\sin z/z\right) $

An interesting situation appears when the energy spectrum of both rings
becomes degenerated. This occurs when the magnetic fluxes threading the
rings are equals ($\phi _{AB}^{u}=\phi _{AB}^{d}=\phi _{AB}$). For this case
we obtain,

\begin{equation}
\mathcal{G}_{\mu }=\frac{e^{2}}{h}\frac{\left( \cos (2\pi \phi _{\mu })-\cos
(z)\right) ^{2}}{\left( \cos (2\pi \phi _{\mu })-\cos (z)\right) ^{2}+4\beta
^{2}}.  \notag
\end{equation}

The spin-dependent conductance vanishes when $\cos (2\pi \phi _{\mu })-\cos
(z)=0$, i.e, when $E_F =\varepsilon _{\mu }^{\alpha }$. The zeroes in the
conductance (Fano antiresonances) represent exactly the superposition of the
spectrum of isolated rings. In fact, the conductance can be written as
superposition of symmetric Fano line-shapes

\begin{equation}
\mathcal{G}_{\mu }=\frac{e^{2}}{h}\frac{\left( \epsilon _{\mu }+q\right) ^{2}%
}{\epsilon _{\mu }^{2}+1}.  \notag
\end{equation}

\noindent where, $\epsilon _{\mu }=\left( \cos (2\pi \phi _{\mu
})-\cos (z)\right) /2\beta $ \ is the detuning parameter and $q$
is the Fano parameter, in this case $q=0$.

\bigskip

\begin{figure}[b!]
\centerline{\includegraphics[angle=0,scale=0.3]{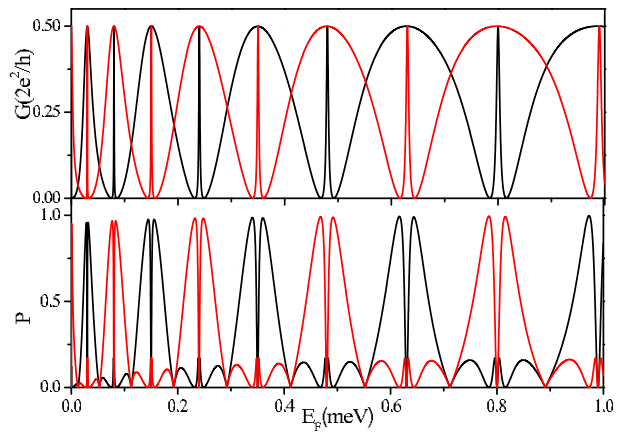}}
\caption{Spin-dependent conductance(upper layer) and spin polarization
(lower layer) as a function of Fermi energy,(color online, black line $%
\protect\mu=+$,red line ($\protect\mu=-$) ) for $\protect\alpha_{SO}=0.5\times 10^{-11}eVm$,$\protect\phi^{u}_{AB}=%
\protect\phi^{d}_{AB}=0.25$.}
\label{fig2}
\end{figure}

Figure 2 displays the spin-dependent linear conductance (upper
layers) and spin polarization (lower layers) versus the Fermi
energy for the symmetric case with $\phi _{AB}=0.25$ and a
spin-orbit coupling $\alpha _{so}=0.5\times 10^{-11}eVm$. The
energy spectrum consists of a superposition of quasi-bound states
reminiscent of the corresponding localized spectrum of the
isolated rings. As expected from the analytical expression (Eq.
\ref{conduc1}) the linear conductance displays a series of
resonances and Fano antiresonances as a function of the Fermi
energy. On the other hand, for given set of parameters the system
shows zones of high polarization due to the splitting of the spin
energy states.

\bigskip

\begin{figure}[b!]
\centerline{\includegraphics[angle=0,scale=0.7]{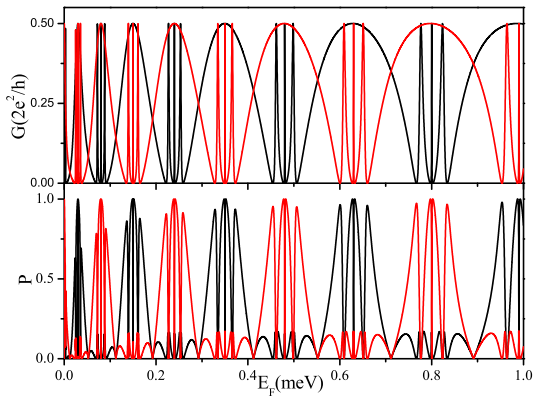}}
\caption{Spin-dependent conductance (upper layer) and spin polarization
(lower layer) as a function of Fermi energy,(color online, black line $%
\protect\mu=+$,red line $\protect\mu=-$) for
,$\protect\alpha_{SO}=0.5\times
10^{-11}eVm$,$\protect\phi^{u}_{AB}=0.3$ and
$\protect\phi^{d}_{AB}=0.2 $} \label{fig3}
\end{figure}

Now we analyze the asymmetric case, i.e $\phi _{AB}^{u}\neq \phi
_{AB}^{d}$. Figure 3 displays the spin-dependent linear
conductance (upper layers) and spin polarization (lower layers)
versus the Fermi energy for a spin-orbit coupling $\alpha
_{so}=0.5\times 10^{-11}eVm$ and parameters of magnetic flux given
by $\phi _{AB}^{u}=0.2,\phi _{AB}^{d}=0.3$. Newly, the zeroes in
the conductance represent exactly the superposition of the
spectrum of each isolated ring $\varepsilon _{\mu n}^{\alpha }$.
In fact, now the conductance vanishes when, $\cos (2\pi \phi _{\mu
}^{u})-\cos (z)=0$ or $\cos (2\pi \phi _{\mu }^{d})-\cos (z)=0,$
i.e when $E_{F}=\varepsilon _{\mu }^{\alpha }$. Notice that due to
the difference between both fluxes new resonances in the
conductance appear. This also affects the structure of the
polarization.

\begin{figure}[b!]
\centerline{\includegraphics[angle=0,scale=0.3]{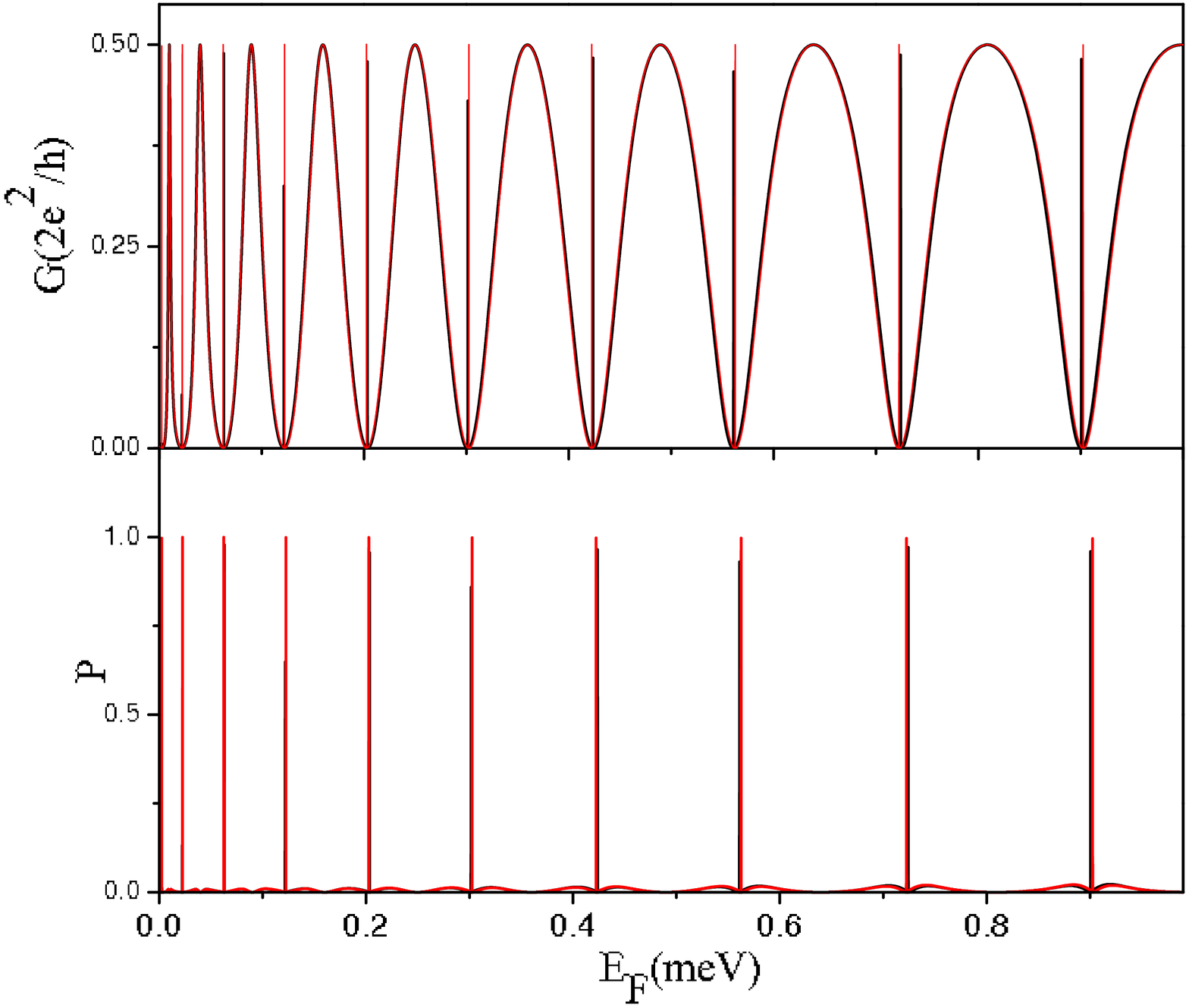}}
\caption{Spin-dependent conductance (upper layer)and spin polarization
(lower layer) as a function of Fermi energy,(color online, black line $%
\protect\mu=+$,red line $\protect\mu=-$) for $\protect\alpha_{so}=0.5\times
10^{-12}eVm$,$\widetilde{\protect\phi}_{AB}=0.25$ and $\protect\delta
\protect\phi _{AB}=0.004988$.}
\label{fig4}
\end{figure}

\begin{figure}[h!]
\centerline{\includegraphics[angle=0,scale=0.3]{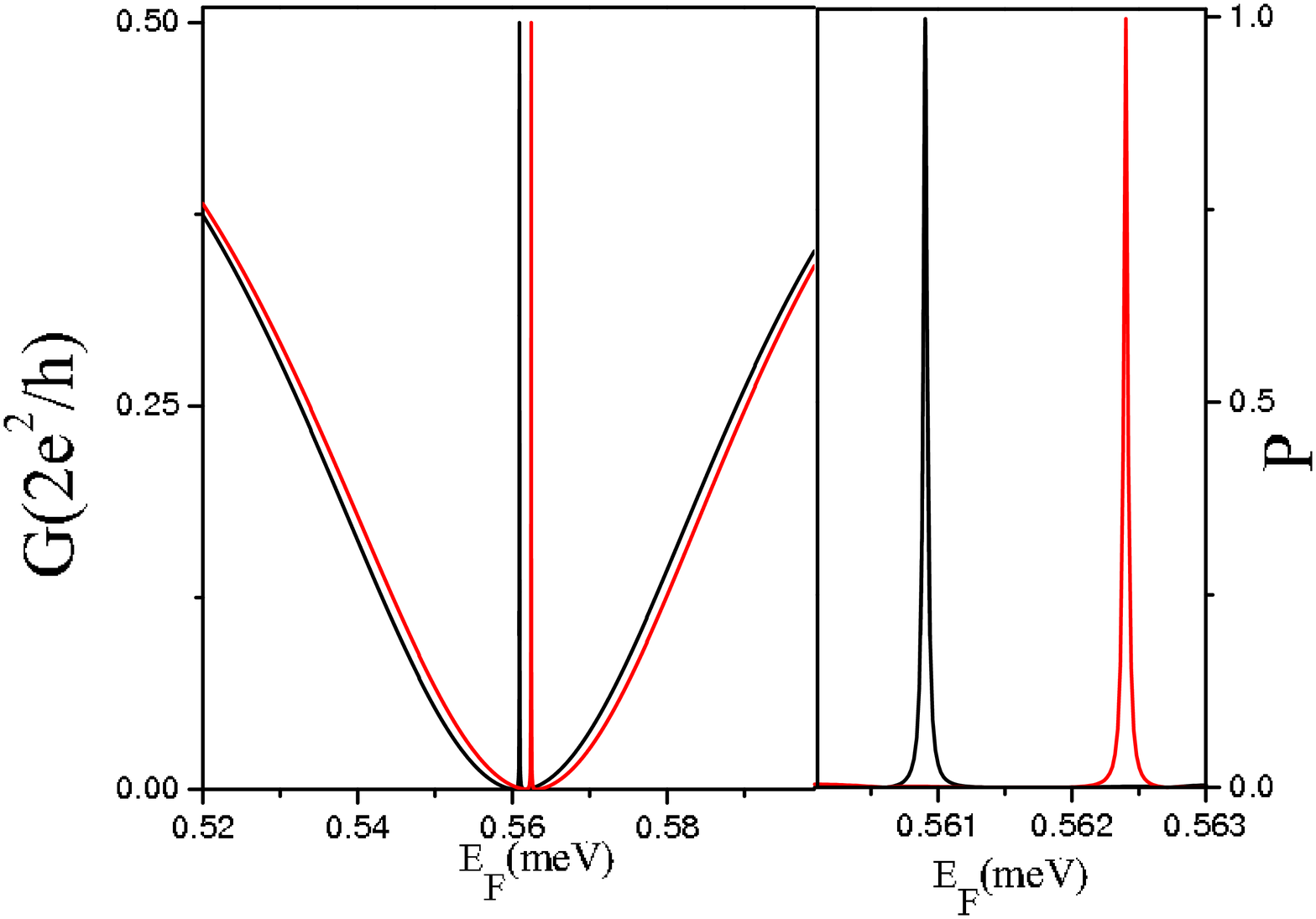}}
\caption{Spin-dependent conductance (left panel) and spin polarization
(right panel) as a function of Fermi energy,(color online, black line $%
\protect\mu=+$,red line $\protect\mu=-$) for $\protect\alpha_{so}=0.5\times
10^{-12}eVm$,$\widetilde{\protect\phi}_{AB}=0.25$ and $\protect\delta
\protect\phi _{AB}=0.004988$.}
\label{fig5}
\end{figure}

\begin{figure}[b!]
\centerline{\includegraphics[angle=0,scale=0.3]{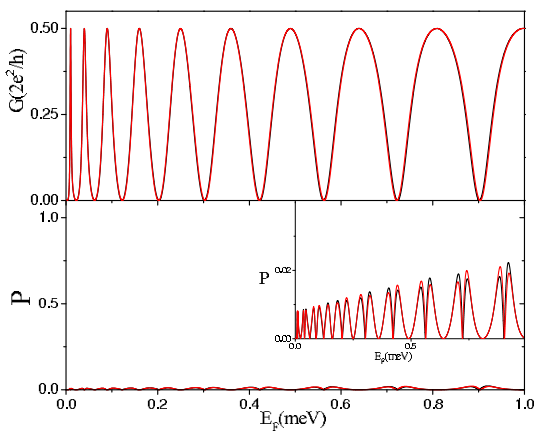}}
\caption{One ring spin-dependent conductance(upper layer) and spin
polarization (lower layer) as a function of Fermi energy,(color online,
black line $\protect\mu=+$,red line $\protect\mu=-$ ) for $\protect\alpha_{SO}=0.5\times 10^{-11}eVm$,$\protect\phi%
_{AB}=0.25$.}
\label{fig6}
\end{figure}

We note that when there is a magnetic flux difference $\delta \phi
_{AB}=\phi _{AB}^{u}-\phi _{AB}^{d}$ high spin polarization can
obtain even for small values of the spin-orbit coupling. In fact,
for small values spin-orbit coupling by adjusting the magnetic
flux difference $\delta \phi _{AB}$ maxima of polarization are
reached. We analyze in detail this situation. The maxima of the
conductance are obtained when $\sin z=0$ or when $\left( \cos
(2\pi \phi _{\mu }^{u})+\cos (2\pi \phi _{\mu }^{d})-2\cos
(z)\right) =0.$ The first condition is spin-independent and it is
not interesting in this case. The second condition is
spin-dependent and for small magnetic flux difference can be
written as, $\cos \left[ 2\pi \left( \frac{\phi _{\mu }^{u}+\phi
_{\mu }^{d}}{2}\right) \right] -\cos (z)\approx 0.$ This occurs
for the energies given by $\widetilde{\varepsilon }_{\mu n}=\hbar \Omega %
\left[ \left( n-\widetilde{\phi }_{\mu }+\frac{1}{2}-\mu \frac{1}{2\cos
\theta }\right) ^{2}-\frac{1}{4}\tan ^{2}\theta \right] ,$where $\widetilde{%
\phi }_{\mu }=\frac{\phi _{\mu }^{u}+\phi _{\mu }^{d}}{2}$ i.e the
position of the maxima of the conductance corresponding to the
spectrum of an effective ring with phase $\widetilde{\phi }_{\mu
}.$ Therefore the condition for the maxima of polarization are
given when the minima of the conductance for one spin-state
coincide with the maxima of the conductance of the opposite spin
(or viceverse), that is $\widetilde{\varepsilon }_{\mu
n}=\varepsilon _{\overline{\mu }n+1}^{\alpha },$ then $\delta \phi
_{AB}=\left( 1-\cos \theta \right) /\cos \theta \approx
\frac{1}{2}\left( \omega _{so}/\Omega \right) ^{2}$. Then, for a
given spin-orbit coupling by adjusting the magnetic flux
difference between the upper and lower rings, the maxima of the
spin polarization are reached. Fig.\ref{fig4} displays the spin
dependent conductance (upper layer) and the spin polarization
(lower layer) for $\widetilde{\phi }_{AB}=0.25$, $\alpha
_{so}=5\times 10^{-12}eVm$ and $\delta \phi _{AB}=0.004988$. The
conductance shows broad and sharp peaks and the spin polarization
shows a series peaks of maximum of polarization. Fig.\ref{fig5}
displays a zoom of the conductance (right panel) and the
polarization (left panel) as a function of the Fermi energy.
Clearly the sharp peaks and Fano antiresonances for the two spin
states are shifted given origin to the peaks of maximum of
polarization. For comparison we plot the corresponding conductance
and polarization for one ring for the same values of the magnetic
flux and spin orbit coupling (Fig. \ref{fig6}). For these
parameter the spin polarization of one ring is very low for both
spin states. The inset in Fig.6 (lower panel) shows a zoom of the
spin polarization.

\bigskip

\bigskip

\begin{figure}[b!]
\centerline{\includegraphics[angle=0,scale=0.3]{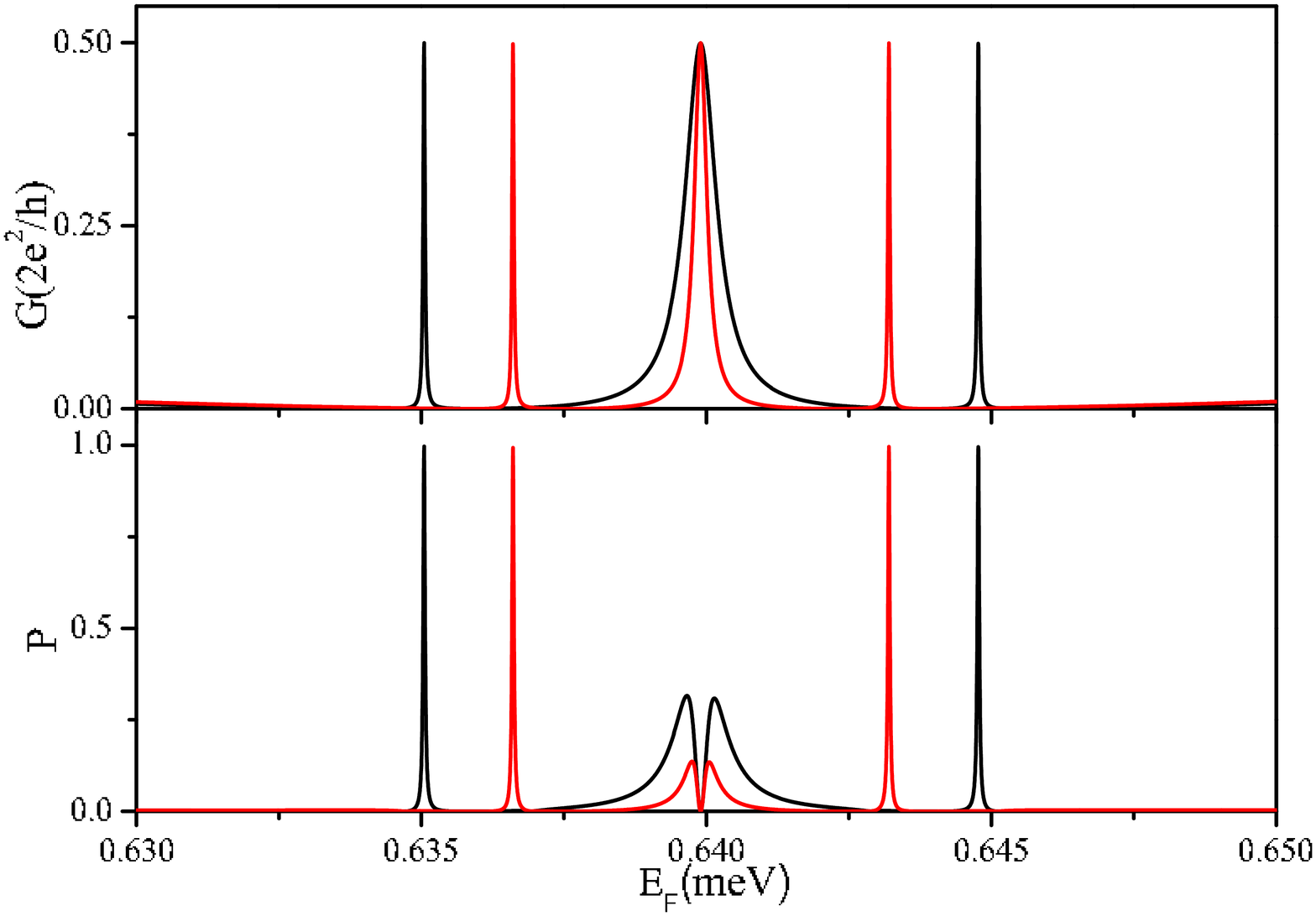}}
\caption{Spin-dependent conductance (upper layer) and spin polarization
(lower layer) as a function of Fermi energy,(color online, black line $%
\protect\mu=+$,red line $\protect\mu=-$ ) for $\protect\alpha_{SO}=0.5\times 10^{-12}eVm$,$\widetilde{\protect\phi}%
_{AB}=0.01$ and $\protect\delta\protect\phi_{AB}=0.004988$.}
\label{fig7}
\end{figure}

For small values of magnetic flux difference $\delta \phi _{AB}$ the
conductance of the two ring system can be written approximately as a
superposition of a broad Fano line shape and a narrow Breit-Wigner line
shape. This is,
\begin{equation}
\mathcal{G}_{\mu }\approx \frac{e^{2}}{h}\left[ \frac{\left( \epsilon _{\mu
}+q\right) ^{2}}{\epsilon _{\mu }^{2}+1}+\frac{\eta _{\mu }^{2}}{x_{\mu
}^{2}+\eta _{\mu }^{2}}\right] .
\end{equation}%
\noindent where the width $\eta _{\mu }=\left( \sin 2\pi \widetilde{\phi }%
_{\mu }\sin 2\pi \delta \phi _{AB}\right) ^{2}/(2\gamma \beta )$
and $x_{\mu }=2\beta \epsilon _{\mu }$. As we discuss in a
previous paper\cite{pedro2}, this expression clearly shows the
superposition of short and long living states developed in the
rings. The apparition of quasi-bound states in the spectrum of the
system is a consequence of the mixing of the levels of both rings
which are coupled indirectly through the continuum of states in
the wire. A similar effect was discussed recently in a system with
a ring coupled to a reservoir by Wunsch et al. in
ref.[\onlinecite{wunsch}]. They relate this kind of collective
states with the Dicke effect in optics. The Dicke effect in optics
takes place in the spontaneous emission of a pair of atoms
radiating a photon with a wave length much larger than the
separation between them. \cite{dicke} The luminescence spectrum is
characterized by a narrow and a broad peak, associated with long
and short-lived states, respectively. This feature allows to
obtain high spin polarization even for small spin-orbit coupling
by adjusting the magnetic flux difference $\delta \phi _{AB}$.
High spin polarization holds even for small values of the magnetic
flux. For instance the Fig. \ref{fig7} displays the conductance
and spin polarization as a function of the Fermi energy for
$\widetilde{\phi }_{AB}=0.01$,$\alpha _{so}=5\times 10^{-12}eVm$)
and $\delta\phi_{AB}=0.004988$. The spin-polarization shows sharp
peaks for the two spin states. As a comparison with a single ring
side-coupled to a quantum wire, the system composed by two rings
allows us to obtain high spin polarization even for small
spin-orbit interaction and small magnetic fluxes, keeping a small
difference for these fluxes.

\bigskip

\section{Summary}

We have investigated the spin dependent conductance and spin
polarization in a system of two side quantum rings attached to a
quantum wire in the presence of magnetic fluxes threading the
rings and Rashba spin-orbit interaction. We show that by using the
Fano and Dicke effects this system can be used as an efficient
spin-filter. We compare the spin-dependent polarization of this
design and the polarization obtained with one ring side coupled to
a quantum ring. As a main result, we find better spin polarization
capabilities as compared to the one ring design. We find that the
spin-polarization dependence for this system is much more
sensitive to magnetic flux and spin-orbit interaction than the
case with only one ring side-coupled to the quantum wire. This
behavior is interesting from theoretical point of view, but also
by its potential technological application.

\section*{ACKNOWLEDGMENTS}

P.\ A.\ O.\ and M.\ P.\ would like to thank financial support from
CONICYT/Programa Bicentenario de Ciencia y Tecnologia (CENAVA, grant ACT27).


\begin{thebibliography}{99}
\bibitem{Chandra} V. Chandrasekhar, R.A. Webb, M.J. Brady, M.B. Ketchen,
W.J. Gallagher and A. Kleinsasser, Phys. Rev. Lett. \textbf{67}
3578 (1991).

\bibitem{Mailly} D. Mailly, C. Chapelier and A. Benoit, Phys. Rev Lett.
\textbf{70} 2020 (1993).

\bibitem{Keyser} U. F. Keyser, C. F\"{u}hner, S. Borck and R. J. Haug, Semm.
Sc. and Tech. \textbf{17}, L22 (2002).

\bibitem{damato} Jorge L. D'Amato, Horario M. Pastawski, and Juan F. Weisz,
Phys. Rev. B textbf{39}, 3554 (1989).

\bibitem{pedro} P.\ A.\ Orellana, M.\ L.\ Ladr\'{o}n de Guevara, M. Pacheco,
and A. Latg\'e, Phys.\ Rev.\ B \textbf{68}, 195321 (2003).

\bibitem{Datta} Datta S and Das B Appl. Phys. Lett. \textbf{56} 665 (1990)

\bibitem{Song} Song J F, Ochiai Y and Bird J P Appl. Phys. Lett. \textbf{82}
4561 (2003)

\bibitem{Folk} Folk J A, R. M. Potok, Marcus C M and Umansky V \emph{%
Science\/} \textbf{299} 679 (2003)

\bibitem{Mireles} Mireles F and Kirczenow G Phys. Rev. B \textbf{64} 024426
(2001)

\bibitem{Mireles2} Mireles F, Ulloa S E, Rojas F and Cota E Appl. Phys.
Lett. \textbf{88} 093118 (2006)

\bibitem{Berciu} Berciu M and Janko B Phys. Rev. Lett. \emph{Lett.}\textbf{90%
} 246804 (2003)

\bibitem{Shelykh} I. A. Shelykh, N. G. Galkin, and N. T. Bagraev Phys. Rev.
B \textbf{74}, 165331 (2006).

\bibitem{bruder} Minchul Lee and C. Bruder Phys. Rev. B \textbf{73}, 085315
(2006).

\bibitem{pedro2} P.\ A.\ Orellana, and M. Pacheco, Phys.\ Rev.\ B \textbf{71}, 235330 (2005).

\bibitem{dicke} R. H. Dicke, Phys. Rev. \textbf{89}, 472 (1953).

\bibitem{Meijer} F. E. Meijer, A. F. Morpurgo, and T. M. Klapwijk, Phys.
Rev. B 66, 033107 (2002).

\bibitem{luo} J. Luo, H. Munekata, F. F. Fang and P. J. Stiles Phys. Rev. B
\textbf{38}, 10142 - 10145 (1988)

\bibitem{wunsch} B. Wunsch, A. Chudnovskiy, Phys. Rev. B \textbf{68}, 245317
(2003).


\end{thebibliography}
\end{document}